\documentstyle[12pt]{article}                     

\def\be{\begin{equation}}
\def\en{\end{equation}}

\begin{document}
\begin{titlepage}
\baselineskip = 25pt
\begin{center} 

{\bf DECONVOLVING THE BEAM IN SMALL ANGULAR SCALE CMB EXPERIMENTS}
 
\vspace{.5 cm}
{\bf J.V. Arnau$^{1}$
and D. S\'aez$^{2}$}\\
\small
$^{1}$Departamento de Matem\'atica Aplicada,
Universidad de Valencia.\\
46100 Burjassot, Valencia, Spain\\
$^{2}$Departamento de Astronom\'{\i}a y Astrof\'{\i}sica, Universidad
de Valencia.\\
46100 Burjassot, Valencia, Spain\\
\footnotesize
e-mail: jose.arnau@uv.es; diego.saez@uv.es\\
\end{center}

\vspace {2. cm}         
\normalsize
\begin{abstract}

This paper is concerned with experiments which measure CMB 
anisotropies on small angular scales. A certain coverage, 
a beam structure and a level of uncorrelated noise define each
experiment. 
We focus our atention on the reversion of the beam average.
In each experiment, 
we look for the best pixelization for reversion, namely, for the 
pixelization that --after reversion-- leads to good maps
containing right spectra for the most wide
range of angular scales.
Squared pixels having different sizes "smaller" than the beam
radius ($\theta_{_{FWHM}}$) are considered.
For a given size, the following question arises:
How well can we assign a temperature to each pixel?
Various mathematical methods are used to show that, in practice, 
this assignation --beam reversion or deconvolution-- 
only leads to right spectra for pixel sizes greater than
a certain lower limit close to $\theta_{_{FWHM}}/2$. 
This limit is estimated for  
negligible  
and relevant levels of noise and also for spherically 
symmetric and asymmetric beams. After this general study, we
focus our attention on 
two feasible detectors (which have been proposed
to be on board of PLANCK satellite). For each of them, we estimate 
the size of the most appropriate pixelization 
compatible with beam reversion, difraction,
observational strategy et cetera and, then,
we answer the following question: Which is 
the part of the angular power spectrum which can be 
extracted from appropriately pixelized maps after deconvolution?

\end{abstract}

{\em Subject headings:} cosmic microwave background---cosmology:theory---
large-scale structure of the universe---methods:numerical

PACS: 98.70.Vc, 95.75.-z, 95.75.pq, 95.75.St

\end{titlepage}
 
\section{INTRODUCTION}                                 

The main goal of this paper is to present a detailed analysis of
the following problem: Given a beam structure, a level of noise, 
a certain partial coverage, and a pixelization, how well can we 
assign a temperature to each pixel?. In other words, how well can we
deconvolve the beam to get appropriate temperatures in the pixels?. 
Here, temperatures are considered to be appropriate when the 
resulting maps lead to good physical spectra. Hereafter, 
any beam reversion leading to 
right spectra is referred to as a "S--deconvolution".
In the absence of noise,
the possibility of performing a good S--deconvolution essentially  
depends on the  
ratio between the beam area and that of the chosen pixel. In practice,
S--deconvolution is not feasible for 
too large values of this ratio; in other words,
if we fix the beam, S--deconvolution is not feasible 
for too small pixel sizes. For a given beam and a certain mathematical
method,
there is a minimum pixel size allowing
S--deconvolution. For values smaller than this 
minimum, too many pixels can be placed inside the beam and 
S--deconvolution is not possible. The minimum size 
corresponding to two S--deconvolution methods
has been 
estimated in various cases. Both methods lead
to similar minimum values of the pixel size around
$\theta_{_{FWHM}}/2$. These values depend on the 
level of the uncorrelated noise produced by the instruments. 

It is worthwhile to emphasize that we are not interested in 
assigning temperatures to hundreds of pixels located inside
the beam. This assignation can be useful in 
other contexts; however, in our case,
the important point is that the spectra 
contained in the S--deconvolved maps must be similar to 
the true physical spectra (up to the scales corresponding to
the pixel size). Unfortunately, this type
of deconvolution requires a moderate number of pixels inside
the beam. This number will be estimated below
in various cases.

From \S \ 2 to \S \ 4,
pixelization is considered in the framework of pure beam reversion,
without analyzing particular experiments;
however, in \S \ 5, we focus our attention 
on PLANCK multifrequency
observations and, then, pixelization is discussed taking into account 
both
previous conclusions about beam reversion and 
some physical constraints due to difraction,
observational strategy, et cetera.

In S\'aez, Holtmann \& Smoot (1996) and S\'aez \& Arnau
(1997), the {\em modified power spectrum} 
\be
E_{\ell}(\sigma)= 
\frac {32 \pi^{3}} {(2 \ell  + 1)^{2}}
\int_{\alpha_{min}}^{\alpha_{max}} C_{\sigma}(\alpha)
P_{\ell}(\cos \alpha) \sin \alpha d \alpha 
\label{modcl}
\en
was described. Functions $P_{\ell}$ are the Legendre polinomial
normalized as follows: $\int P_{\ell} P_{\ell^{\prime}} d(\cos \theta)
= [(2 \ell + 1)/8 \pi^{2}] \delta_{\ell \ell^{\prime}}$.
As explained in those papers, this type of spectrum 
can be easily found from both theory and maps. Comparisons of the
modified spectra obtained from theory with those extracted from simulated
or observational maps are appropriate to take into account
pixelization, partial coverage and beam features, simultaneously. 
Sometimes, the estimation and use of the well known $C_{\ell}$
coefficients --although possible-- 
is not the best procedure. In Eq. (\ref{modcl}),
the effect of pixelization is simulated by the angle $\alpha_{min}$, 
which is the angle separating two neighbouring nodes, while the
angle $\alpha_{max}$ depends on the area of the covered region;
this angle is to be experimentally obtained
(S\'aez \& Arnau, 1997), it is smaller than the
size of the map and large enough to include as much scales as  
possible. The autocorrelation function $C_{\sigma}(\alpha)$ is
\be 
C_{\sigma}(\alpha)= \left< \left( \frac {\delta T}{T} \right)_{\sigma}
(\vec {n}_{1}) \left( \frac {\delta T}{T} \right)_{\sigma}
(\vec {n}_{2})
\right>   \ ,
\label{cete}
\en
where $\alpha$ is the angle formed by the unit vectors $\vec {n}_{1}$ and
$\vec {n}_{2}$, the angular brackets stand for an average over many full 
realizations of the CMB sky and, quantity $( \delta T/T)_{\sigma} (\vec {n})$
is the temperature contrast in the direction $\vec {n}$ after 
smoothing with a Gaussian beam having a certain 
$\sigma = 0.425 \theta_{_{FWHM}}$.
The modified spectra are used 
below to analyze some simulated maps.

\section{SIMULATIONS}  

The angular power spectrum 
$C_{\ell} = \sum_{m=- \ell}^{m= \ell} |a_{\ell m}|^{2}/(2 \ell +1)$
is only
an auxiliary element in our estimations. We are not
particularly interested in any choice and, consequently, 
we have used the same 
spectrum as in S\'aez, Holtmann \& Smoot (1996). It          
corresponds to the minimum 
cold dark matter model with a
baryonic density parameter $\Omega_{_{B}}=0.03$ and
a reduced Hubble constant $h=0.5$. 
The $C_{\ell}$ coefficients 
have been taken from Sugiyama (1995) and renormalized according to
the four-year COBE data ($Q_{rms_PS} \simeq 18 \mu K$ , Gorski et al.
1996). Our simulations are performed 
by using the Fast Fourier Transform (see S\'aez, Holtmann \& Smoot 1996
and Bond \& Efstathiou 1987) and, then, a certain beam is used to
average temperatures;
thus, we obtain maps which must be deconvolved with the same beam.
After S--deconvolution, the resulting map must be compared with 
the initial one.

\section{IDEAL BEAM S--DECONVOLUTION}

Two methods are proposed to perform a S--deconvolution of the beam
in the absence of noise (ideal case).
The efficiency of these methods is verified and
the limits for their application are discussed.
For appropriate coverages and beam structures,
the size of the smallest pixels compatible with 
beam reversion is estimated in each case.
The conclusions obtained in this section are important 
to understand realistic S--deconvolution in noisy maps (\S \ 4).

\subsection{BEAM}

We begin with a Gaussian spherically symmetric beam. 
If the direction of the beam center
is $\vec {n}$, the measured temperature $T(\vec {n})$ is given by the 
following average:
\be
T(\vec {n})=\frac {1}{2 \pi \sigma^{2}} \int 
T^{*} (\theta , \phi) e^{-[\theta^{*}
(\vec {n})]^{2}/2
\sigma^{2}} \sin \theta d \theta d \phi   \ ,
\label{beam0}
\en    
where $\sigma$ defines the beam size, the angles      
$\theta$ and $\phi$ are the spherical coordinates
of a certain pixel, the element of 
solid angle is 
$d \Omega = \sin \theta d \theta d \phi$ and,
quantity $\theta^{*}$
is the angle formed by the direction ($\theta$, $\phi$) and the observation 
direction $\vec {n}$. 

Small pixels can be considered as surface 
elements and, consequently, Eq. (\ref{beam0}) can be discretized as
follows:
\be
T(\vec {n})=\frac {1}{2 \pi \sigma^{2}} \sum_{i} 
T_{i}^{*} e^{-(\theta_{i}^{*})^{2}/2 \sigma^{2}} 
\frac {dS_{i}^{2}}{R^{2}} \ ,
\label{beam1}
\en    
where the subscript $i$ stands for the $i$-th pixel; here, $T^{*}_{i}$
and $dS_{i}$ are the temperature and the area of the $i$-th pixel,
respectively, and $\theta^{*}_{i}$ is the angle formed by this pixel and
the beam center. The exponential tends rapidly to zero as $\theta_{i}^{*}$
increases beyond $\sigma$; hence, only a small number of pixels are 
significant in order to estimate $T(\vec {n})$. 
Furthermore, due to technical reasons,
the beam could receive energy from a reduced number of pixels
(not from all the significant pixels in an ideal infinite Gaussian
beam). By these reasons, we
assume that only $q \times q$
pixels are relevant and we give various values 
to number $q$; these pixels cover a square patch 
centered at the same point as the beam.

An asymmetric beam of the form
\be
W= \frac {1}{2 \pi \sigma^{2}} e^{[a^{2}(\theta - \theta^{\prime})^{2}
+a^{-2}(\phi - \phi^{\prime})^{2}]/2 \sigma^{2}}
\label{beam2}
\en
has been also considered. The parameter $a$ defines the degree of 
asymmetry. 
                                
Figure 1 illustrates, for $q=7$, three situations corresponding
to beams and pixelizations
considered below. 

\subsection{COVERAGE}

Our choice of an appropriate partial coverage is based on
some results obtained in  
previous papers.
In S\'aez, Holtmann \& Smoot (1996), it was shown that CMB maps 
close to $20^{\circ} \times 20^{\circ}$ can be simulated --with good 
accuracy-- neglecting curvature and using the
Fourier transform. 
The effects of partial coverage were
analyzed in detail in S\'aez \& Arnau (1997), these authors     
proved that, although 
a $20^{\circ} \times 20^{\circ}$ map does not suffice to get the
angular power spectrum, a few tens of $20^{\circ} \times 20^{\circ}$ 
maps allow us to find the power spectrum 
with a high accuracy for $\ell > 200$. On account of this previous result,
forty $20^{\circ} \times 20^{\circ}$ maps are used in this paper to get 
each angular power spectrum. These maps would
cover about $40 \%$ of the sky.
In the above papers, it was also argued that uncorrelated noise 
is not expected to be important for coverages
greater than or equal to $20^{\circ} \times 20^{\circ}$ and for the noise
level expected in modern CMB experiments. 
This is true in the sense that the power spectrum can be calculated in the
presence of this noise, but the noise can be problematic 
for S--deconvolution as we will discuss below.
These comments point out
the interest of considering $20^{\circ} \times 20^{\circ}$ maps 
and motivate our choice 
of these regions to begin with our analysis 
of the S--deconvolution procedure.

\subsection{PIXELIZATION AND EQUATIONS TO BE SOLVED} 

In order to compute the integral in (\ref{beam0}) using Eq. (\ref{beam1}),
the area $dS_{i}$ is not required to be independent on $i$; namely,
no equal area distributions of pixels are necessary. Furthermore,
if we take a large enough number of small pixels covering all 
the region contributing significantly to the integral
(\ref{beam0}), moderated variations in the pixel shapes
are also admissible. In spite of these comments,
equal area and equal shape 
pixelizations are advantageous --at least from the
mathematical point of view-- as it is shown along the paper. 

Since a $20^{\circ} \times 20^{\circ}$ region is approximately
flat, small squares with edges of angular lenght $\Delta \theta =
\Delta \phi = \Delta$ define an approximately 
regular pixelization. The number of pixels per edge is $N=20/ \Delta$,
where angle $\Delta$ must be given in degrees.
Our measurements
would cover the pixelized region with 
a certain strategy. The beam center should point towards each pixel $\alpha$
various times and, then, the mean of the resulting measures could be 
considered as the smoothed temperature 
$T_{\alpha}$ corresponding to pixel $\alpha$. From these $T_{\alpha}$
values, a new temperature which does not
involve the beam effect, $T^{*}_{i}$, must be assigned to each pixel
$i$; namely,
the beam must be deconvolved. All along \S \ 3, 
it is assumed that sistematic errors have been corrected and 
that the uncorrelated instrument noise is negligible. If $T_{\alpha}$
is the temperature measured by the instrument when the beam center is 
pointing towards the center of the pixel $\alpha$, according to Eq.
(\ref{beam1}), we can write
\be
T_{\alpha } = \sum_{i} A_{\alpha i} T^{*}_{i}
\label{sis}
\en
with
\be
A_{\alpha i} = \frac {1}{2 \pi \sigma^{2}} 
e^{-(\theta_{i}(\vec {n}_{\alpha})^{*})^{2}/2 \sigma^{2}} 
\frac {dS_{i}^{2}}{R^{2}}  \ ,
\label{cosis}
\en
where $T^{*}_{i}$ is the true temperature in the i-th pixel; namely,
the S--deconvolved temperature we are looking for. Similar equations hold
for the asymmteric beam (\ref{beam2}).

It is evident that the temperatures $T^{*}_{i}$
only define a S--deconvolution if they are     
an approximate solution of the
linear Eqs. (\ref{sis}). Only in this case, 
the temperatures $T^{*}_{i}$ are similar to the 
true temperatures averaged by the beam and, consequently, 
the resulting 
maps contain the right spectra. This fact strongly restrict
the methods for S--deconvolution.
Two of them are described in next section.
These
equations
can be written in the matrix form $T = A T^{*}$,
where $T$ and $T^{*}$ are arrays of $N \times N$ numbers (one number
for each pixel) and A is a 
$N^{2} \times N^{2}$ matrix. The element $A_{\alpha i}$
weights the contribution of the pixel $i$ to the smoothed
temperature at pixel $\alpha$. Quantity $A_{\alpha i}$ is 
assumed to be significant only
in the $q \times q$ pixels mentioned in \S \ 3.1.

\subsection{METHODS AND RESULTS} 

Two methods are used to estimate the S--deconvolved 
temperature $T^{*}_{i}$ corresponding
to pixel $i$: In the first one, Eqs. (\ref{sis}) are solved as a linear 
system of  algebraic 
equations (hereafter LS--deconvolution) where       
the independent terms are 
the observed temperatures  $T_{\alpha}$.
In the second method,
equation (\ref{beam0}) is considered as a convolution and, then,
Fourier transform (FT) and the deconvolution
theorem are used to get $T^{*}_{i}$ on the nodes of the 2D Fourier 
grid (hereafter FTS--deconvolution).

\subsubsection{LS--DECONVOLUTION}

Iterative methods (Golub \& van Loan, 1989; 
Young, 1971) can be used to solve the system
(\ref{sis}). In any of these methods, the matrix $A$ is split
as follows A=M-P, where M is any matrix which can be easily inverted.
In matrix form, the iteration scheme reads as follows:
\be
T^{*(n+1)}=M^{-1}PT^{*(n)}+M^{-1}T  \ ,
\label{iter1}
\en
where the superscript $n$ stands for the n-th iteration. The necessary 
and sufficient 
condition for convergence is that the spectral radius  of the 
matrix $Q=M^{-1}P$ is smaller than unity. This radius is 
the maximum 
of $| \lambda_{i} |$, where $ \lambda_{i} $ are the 
eigenvalues of $Q$.       
Hereafter, the so-called Jacobi method is used. This method
corresponds to a particular choice of the matrix $M$. This matrix is 
assumed to be the matrix formed by the diagonal of $A$, which is 
denoted $A_{_{D}}$. A sufficient condition for the convergence of the  
Jacobi method is that matrix $A$ is diagonal dominant
($|A_{ii}|> \sum_{j \neq i} |A_{ij}|$ for any $i$).
The dimension of the 
matrices $A$, $A_{_{D}}$ and $P$ are $N^{2} \times N^{2}$, where N is
the number of pixels per edge in the map. Since this number is
greater than $10^{2}$ in all the practical cases, the dimension
of the above matrices is very great and they cannot
be stored. Fortunately, this
storage is not necessary; in fact, if Eq. (\ref{iter1}) is rewritten 
using indices
\be 
A_{ii}T^{*(n+1)}_{i}=T_{i} - \sum_{j<i} A_{ij}T^{*(n)}_{j}
- \sum_{j>i} A_{ij}T^{*(n)}_{j} \ ,
\label{iter2}
\en
we see that all the elements of the matrix $A$ 
necessary to get $T^{*(n+1)}_{i}$
can be obtained when they are necessary, without storage.
Of course, a given element can be calculated various times,
but no storage is necessary at all.

Equation (\ref{iter1}) can be considered as a matrix equation in
which each element is a block. Matrix $A$ is split in $N \times N$
blocks and vectors $T$ and $T^{*}$ in N arrays (blocks) of dimension
$N$. Then the resulting $N \times N$ 
blocks appear to have a q-diagonal structure and 
diagonal domination 
gives limit values for quantities $\lambda =
e^-(\Delta^{2}/2 \sigma^{2})$ and $\theta_{_{FWHM}}/ \Delta$.
The maximum value of $\lambda$ and the corresponding maximum value 
of the ratio $\theta_{_{FWHM}}/ \Delta$
are given in the first and second
columns of Table 1, respectively, for various $q$ values. The third column 
gives the ratio
$S_{q}/S_{_{B}}$, where $S_{q}$ is the area covered by the
$q \times q$ significant pixels and $S_{_{B}}$ is the area of the beam
(a circle with radius $\theta_{_{FWHM}}$). The top 
panel of Fig. 1 illustrates the case $q=7$. Eleven pixels are located
inside the beam circle (radius $\theta_{_{FWHM}}$). 

Taking into account that, in general,
the condition $\lambda < \lambda_{max}$ is only a sufficient 
condition for the convergence of the Jacobi method (not necessary),
we have studied numerically many cases in which 
$\lambda \geq \lambda_{max}$ 
with the hope of getting new convergent cases.
The result is that the
Jacobi method has never converged for $\lambda > \lambda_{max}$. 
This result points out that, in practice, 
the condition $\lambda \leq \lambda_{max}$
is also necessary for LS--deconvolution. This information 
suffices for us. A rigorous mathematical
study about the necessary character of this
condition is not appropriate here. 
From the intuitive point of view, 
the existence --in practice-- of 
a $\lambda_{max}$ value is an expected result, in fact, 
in the absence of a $\lambda_{max}$, 
it would be possible to assign deconvolved temperatures
to billions (an arbitrary number)
of small pixels placed inside the beam and, furthermore,  
the right spectrum could be recovered up to the spatial scales
of these small pixels; this would be a nonsense.

When LS--deconvolution applies, it is an accurate 
method. In fact, for $\theta_{_{FWHM}} =8.8^{\prime}$ and 
$\Delta=4.6875^{\prime}$, 
about 24 iterations suffice to get a very good 
deconvolved map. After these iterations, the relation
$[\sum_{i=1}^{N^{2}} (T^{(n+1)}_{i} - T^{(n)}_{i})^ {2}]^{1/2}/N^{2}<
10^{-7}$ is satisfied and, consequently, the method  
is converging towards a certain map. The question is: are the
numerical iterations converging to a good S--deconvolved
map with the right spectrum? In order to answer this question,
we proceed as follows:
(1) the Fast Fourier Transform (FFT) is used to do a
simulation ($S_{1}$) which does not involve either beam or noise,
(2) a second map ($S_{2}$) is obtained --from $S_{1}$-- by using 
a certain beam for smoothing,
(3) the map $S_{2}$ is deconvolved using the Jacobi
method to get a 
new map ($S_{3}$), (4) the three above steps are repeated  
forty times (see \S \ 3.2) and, (5) the method 
described in the introduction (see also
S\'aez, Holtmann \& Smoot 1996 and S\'aez \& Arnau
1997) is used
to obtain the modified spectrum in three cases:
before smoothing (from $S_{1}$ maps),
after smoothing (from $S_{2}$ maps) and, after deconvolution 
(from $S_{3}$ maps), these spectra are hereafter referred to
as $E_{1 \ell }$, $E_{2 \ell }$, and $E_{3 \ell }$, respectively. 
If deconvolution is a
good enough S--deconvolution, spectrum $E_{3 \ell }$ should be 
comparable with
$E_{1 \ell }$. Whatever the deconvolution method 
may be,          
these five steps allow us to analize the resulting deconvolved maps.
In all the Figures of this paper which show the three above spectra,
pointed, dashed, and solid lines 
correspond to $E_{1 \ell }$, $E_{2 \ell }$, and $E_{3 \ell }$, 
respectively.
The top right panel of Fig. 2 shows 
the resulting spectra for the LS--deconvolution under
consideration ($\theta_{_{FWHM}} =8.8^{\prime}$ and $\Delta=
4.6875^{\prime}$). 
We see that the dotted line ($E_{1 \ell }$)
is almost indistinguishable from the solid one ($E_{3 \ell }$)
for $\ell < 2000$. This result qualitatively 
proves the goodness of the iterative LS--deconvolution.
In order to compare the spectra $E_{1 \ell }$ and 
$E_{3 \ell }$ quantitatively, the following 
quantities are calculated and presented in Table 2 (entries 7 and 8):
The mean, $M1$, of the quantities $\ell ( \ell + 1 ) E_{\ell} \times
10^{10}$ corresponding to the spectrum $E_{1 \ell }$ (col. [3]), the mean, 
$M2$, of the differences $E_{1 \ell }-E_{3 \ell }$  
(col. [4]), the mean MA of $| E_{1 \ell }-E_{3 \ell } |$
(col. [5]), and the typical deviation, $\Sigma$, of the differences of 
column (4) (col. [6]). The above quantities are computed in appropriate 
$\ell$--intervals (col. [2]). 
Entries 7 and 8 show values of $|M2|$, $MA$ and $\Sigma$ much smaller than 
$|M1|$, which means that the spectra $E_{1 \ell }$ and 
$E_{3 \ell }$ are very similar in both $\ell$--intervals (40,1000) and
(1000,2000). It is also remarkable that $|M2|$ is much smaller than 
MA, which means that spectrum 
$E_{3 \ell }$ oscillates around spectrum $E_{1 \ell }$ 
giving positive and negative values of $E_{1 \ell }-E_{3 \ell }$
which cancel among them.

Table 2 compares other pairs of spectra
displayed in Figs. 2 and 3.  Column (1) gives the 
Figure and panel where each pair of spectra are displayed.
Tables 3 and 4 have the same structure as Table 2, but they compare  
pairs of spectra contained in Figs 4 and 5, respectively.
The interpretation of the data exhibited in these Tables is 
straightforward. For a given entry, the compared spectra are
similar if the quantities $|M2|$, $MA$ and $\Sigma$ are
much smaller than $|M1|$. Given two entries with similar 
$|M1|$ values, the smaller the values of 
$|M2|$, $MA$ and $\Sigma$, the greater the similarity between the
spectra (the better the deconvolution if we are comparing 
$E_{1 \ell }$ and $E_{3 \ell }$ spectra).

\subsubsection{FTS--DECONVOLUTION}

The method based on the FT only can be used in the case of
small enough coverages (almost flat regions)
allowing a uniform pixelization.                               
The region covered by the observations should be 
a square with no much more than $\sim 20^{\circ}$ per edge; thus, curvature 
can be neglected and the covered area can be considered as 
a square where the angles $\theta$ and $\phi$ play the role of
cartesian coordinates.
Equation (\ref{beam0}) can be then seen as a convolution of the function
$U(\theta,\phi) = T^{*} (\theta,\phi) \sin (\theta)$ with the 
Gaussian beam function 
$W= \frac {1}{2 \pi \sigma^{2}} e^{[(\theta - \theta^{\prime})^{2}
+(\phi - \phi^{\prime})^{2}]/2 \sigma^{2}}$, where coordinates
$\theta^{\prime}$ and $\phi^{\prime}$ define the observation direction
$\vec {n}$. Then, the deconvolution theorem ensures that
the Fourier transform of function $U$ is
\be
U(\vec {k})=T(\vec {k})/W(\vec {k}) \ ,
\label{conv}
\en 
where $\vec {k}$ is a vector in the 2-dimensional Fourier space.
Given a smoothed map and a window function, we can find their Fourier
transforms $T(\vec {k})$ and $W(\vec {k})$ and, 
then, Eq. (\ref{conv}) plus an inverse FT allows us to find the
deconvolved map.
Unfortunately, the use of the FT is not compatible with 
spherically assymmetric rotating beams. If one of these beams 
measures in such a way that its orientation changes from measure 
to measure, Eq. (\ref{beam0}) is not a convolution anymore and,
consequently, the FTS--deconvolution does not apply;
hence, the FT can be used either in the case of a 
spherically symmetric beam or in the case of a
nonspherical nonrotating beam which measures preserving 
its orientation.
 
The maximum value of the ratio $\theta_{_{FWHM}}/ \Delta$
compatible with FTS--deconvolution has been derived using simulations.
In all the $20^{\circ} \times 20^{\circ}$ simulations, we have taken 
$N=256$ ($\Delta=4.6875^{\prime}$), while quantity 
$\theta_{_{FWHM}}$ has been varied appropriately.
The code for FTS--deconvolution has been 
run in each case. This code follows  
the five steps of the process described above for 
analyzing deconvolved maps.
The left panels of Fig. 2 show the spectra 
$E_{1 \ell }$, $E_{2 \ell }$, and $E_{3 \ell }$, 
for
different values of $\theta_{_{FWHM}}$. The top left panel, which 
corresponds
to $\theta_{_{FWHM}}=10^{\prime}$, shows that the 
spectra $E_{1 \ell }$ and $E_{3 \ell }$ are very similar in all the 
$\ell $--interval (2,2000). For $\theta_{_{FWHM}} \sim 11^{\prime}$
(middle left panel), these
spectra are very similar for $2 \leq \ell \leq 1000$, while 
they become a little diferent in the interval (1000,2000). 
For $\theta_{_{FWHM}}>11^{\prime}$ , the differences between 
$E_{1 \ell }$ and $E_{3 \ell }$ grow rapidly 
and, for $\theta_{_{FWHM}}=11.5^{\prime}$ (bottom left panel), 
these spectra are very different. 
This is quantitatively confirmed by the numbers presented 
in entries 1 to 6 of Table 2, where we see that, in the interval
(1000,2000), quantities $|M1|$, $MA$ and $\Sigma$ increase 
as $\theta_{_{FWHM}}$ does.
The maximum value of $\theta_{_{FWHM}}/ \Delta$ appears to be
$\sim 2.3$. This means that, in the case 
$1.87 < \theta_{_{FWHM}}/ \Delta \leq 2.3$, the 
FTS--deconvolution applies and the
LS--deconvolution does not. The most dense grid compatible with 
FTS--deconvolution is shown in the middle panel of Fig. 1. In this case, 
around 16 pixels can be placed inside the beam circle.             

An asymmetric nonrotating beam of the form (\ref{beam2})
has been also deconvolved for various values of the parameters 
$a$ and $\theta_{_{FWHM}}$. 
The middle right panel of Fig. 2 shows the results 
of the FTS--deconvolution for
$a=1.29$ and $\theta_{_{FWHM}}=7.75^{\prime}$. 
These results are good in the full $\ell$--interval (2,2000)
(see entries 9 and 10 of Table 2).
This choice of
the parameters $a$ and $\sigma$ simulates an asymmetric beam whose
effective $\theta_{_{FWHM}}$ along the $\theta$-axis ($\phi$-axis) is
$\theta^{eff}_{1} = \theta_{_{FWHM}}/a \simeq 6^{\prime}$
($\theta^{eff}_{2} = a \theta_{_{FWHM}} \simeq 10^{\prime}$).
This beam and the most dense pixelization allowing its
FTS--deconvolution are shown in the bottom panel of Fig. 1.
Around nine pixels are located inside the beam ellipse.
Other $a$ and $\sigma$ values have been also considered to conclude
that FTS--deconvolution is possible when both
$\sigma^{eff}_{1}/ \Delta$ and 
$\sigma^{eff}_{2}/ \Delta$ are smaller than $\sim 2.3$ (this
constraint is equivalent to that obtained in the spherically 
symmetric case). 
Finally,
we have compared the effect of 
a symmetric beam with $\theta_{_{FWHM}} = 10^{\prime}$ 
and that of the asymmetric nonrotating beam described above. 
In order to do 
this comparison we have obtained forty $S2$ maps with each beam.
The spectra 
obtained from these S2 maps are displayed in the
bottom right panel of Fig. 2 and quantitatively compared in
entries 11 and 12 of Table 2. The solid (pointed)
line corresponds to the asymmetric (symmetric) beam. 
Results show that the deformations 
of the original spectrum produced by these beams 
are different; namely, that the $E_{2 \ell}$
spectra are distinct (significant asymmetry). 

Vanishing instrumental noise has been assumed so far; nevertheles,
partial coverage introduces a kind of sky noise
(an uncertainty). In order to estimate
this noise for a coverage of forty $20^{\circ} \times 20^{\circ}$
maps, we compare the $S1$ spectrum extracted from these maps 
with the theoretical spectrum (which would correspond to many realizations 
of the full sky).
Both spectra are presented in Fig. 3, where the solid (pointed) line 
corresponds to the S1 (theoretical) spectrum. The quantitative 
comparison of these spectra is given in entries 13 and 14 of Table 2.
We see that the deviations with respect to the true spectrum --produced by
the partial coverage under consideration-- are greater than those produced by 
good S-deconvolutions (compare entries 13 and 14 of Table 2 with 
the pairs of entries 1--2, 7--8, and 9--10. Compare also 
the corresponding panels
in the Figures). The deviations decrease as the coverage increases.

\section{BEAM DECONVOLUTION IN NOISY MAPS}

In \S \ 3, negligible uncorrelated noise has been assumed,
thus, for a given beam, a minimum pixel size
for LS--deconvolution and another one for FTS--deconvolution
have been found. These minima define
theoretical restrictions for admissible pixelization; nevertheless,
in the presence of uncorrelated noise, 
stronger restrictions on pixel sizes could appear and,
consequently, S--deconvolution
could be impossible for some sizes close to the
minimum size obtained in the absence of noise. Which is the 
minimum size allowing FTS--deconvolution in the presence of 
a certain level of uncorrelated noise?
We are going to study this question.

As it is well known, the amount of noise in a map depends 
on the observing time per pixel, $t_{pix}$, which is inversely proportional 
to the pixel area. At pixel i, the noise contributes to the temperature
an amount $\delta T^{^{N}}_{i}$. 
It is assumed that the noise is uncorrelated and has uniform variance
$\sigma_{_{N}}^{2}$; i.e., $\langle  \delta T^{^{N}}_{i} 
\delta T^{^{N}}_{j} \rangle =
\sigma_{_{N}}^{2} \delta_{ij}$. The relation 
$\sigma_{_{N}} = s/(t_{pix})^{1/2}$
can be used to estimate the level of uncorrelated noise in 
the pixelized map, where $s$ is the detector sensitivity (see Knox, 1995).
For $s= 200 \ \mu K \sqrt (sec)$ , $\Delta=4.6875^{\prime}$, and                      
a year of uniform full-sky coverage, the level of noise is
$\sigma_{_{N}} \simeq 93.9 \ \mu K$. Furthermore,
using    
the $C_{\ell}$ numbers of \S \ 2 with 
$\Delta=4.6875^{\prime}$ and $\theta_{_{FWHM}}=8.8^{\prime}$,
the expected signal $S=[(1/ 4 \pi) ( \sum _{\ell} (2 \ell + 1)
C_{\ell} e^{- \sigma^{2} \ell^{2}})]^{1/2}$ 
takes on the value $S \simeq 108 \mu K$. Therefore,
for the above choice of $s$, $\theta_{_{FWHM}}$,
and $\Delta$, the signal to noise ratio $S/ \sigma_{_{N}}$
is close to $1$ and, consequently, 
noise cannot be neglected {\it a priori}
in order to do beam S--deconvolution. 
In other realistic cases, the situation is similar.
Since $\sigma_{_{N}}$
is proportional to $(t_{pix})^{-1/2}$, too long observation times 
would be necessary to rise 
$S/ \sigma_{_{N}}$ significantly. 
Fortunately, technological progress leads to
smaller and smaller
$s$ values and, accordingly, the ratio $S/ \sigma_{_{N}}$
increases.
In the PLANCK project of the European Spatial Agency there are two
instruments: the Low Frequency Instrument (using radiometers) and     
the High Frequency Instrument (using bolometers).  
One of the radiometers
($\nu=100 GHz$ and $\theta_{FWHM}=10^{\prime}$),
working at $20^{\circ} \ K$, would produce a noise 
$\sigma_{_{N}} \sim 54 \ \mu K$ (for pixels with 
$\Delta=4.6875^{\prime}$)
during a year of uniform coverage; moreover,  
for one of the bolometers 
($\nu=143 GHz$ and $\theta_{FWHM}=10.3^{\prime}$),
which would work at $\sim 0.1^{\circ} \ K$, the noise would be  
$\sigma_{_{N}} \sim 16.4 \ \mu K$ 
for the same pixels and time coverage.
These data are taken into account below in order to analyze the 
perspectives of beam S--deconvolution in the framework of 
the most accurate project for anisotropy detection 
in small angular scales 
(the PLANCK mission of the European Spatial Agency).

In spite of the fact that LS--deconvolution has been very useful 
in order to analyze and understand the existence of 
a minimum size for pixelization, in practice, 
only the FTS--deconvolution has been used --so far--
in the noisy case.
The maximum value of $\theta_{_{FWHM}}/ \Delta$ compatible with 
FTS--deconvolution depends on the level of uncorrelated noise.
For this type of noise, 
if the average in Eq. (\ref{cete}) is performed on many
sky realizations, the resulting $C_{\sigma}( \alpha )$
values must be very small; nevertheless, if the average is
done in a 
$20^{\circ} \times 20^{\circ}$ patch, the
$C_{\sigma}( \alpha )$ values can be relevant, which means
that, on the patch, the noise is not properly 
uncorrelated (its spectrum is unknown). 
This fact is important in order to understand
the effect of this noise on FTS--deconvolution
of $20^{\circ} \times 20^{\circ}$ maps. 
In the presence of a certain noise which is independent on the
signal, FTS--deconvolution could be performed using the so-called
optimal Wiener filter (Press et al. 1988). In such a case,
the spectrum of the function $U$ --defined above-- 
is estimated as follows:
\be
U(\vec {k})=T(\vec {k}) \Phi (\vec {k}) / W(\vec {k}) \ ,
\label{convo}
\en 
where 
\be
\Phi (\vec {k})= 1 - \frac {|N(\vec {k})|^{2}}
{|U(\vec {k})|^{2} + |N(\vec {k})|^{2}} \ .
\label{filtro}
\en
In the absence of noise, function $\Phi$ takes on the
form $\Phi (\vec {k})= 1$ and Eq. (\ref{convo}) reduces to 
Eq. (\ref{conv}).
Equations (\ref{convo}) and (\ref{filtro}) cannot be 
used in practice to deconvolve the beam (unknown 
spectrum of a given $20^{\circ} \times 20^{\circ}$ noise
realization); nevertheless, these equations 
are useful to understand why the noise can be neglected in 
some cases.
The maximum of the $|U(\vec {k})|^{2}$ values corresponding to
forty $20^{\circ} \times 20^{\circ}$ simulations --based on the
model of \S \ 2-- has been estimated to be $1.7$, while
the maximum of $|N(\vec {k})|^{2}$ obtained from the same number of
simulations of 
pure noise (uncorrelated in great regions) has appeared to be
proportional to the level of noise $\sigma_{_{N}}$. For 
$\sigma_{_{N}} = 16.4 \ \mu K$, the resulting maximum is
$2.7 \times 10^{-3}$. In this case --and also
for any current or planned experiment--   
the amplitude corresponding to $|N(\vec {k})|^{2}$
is much smaller than that of $|U(\vec {k})|^{2}$. This 
smallness
--relative to that of the signal-- indicates that, in realistic
noisy cases,
the filter function is close to $\Phi (\vec {k})= 1$ and
the following question arises: Is it possible to take
$\Phi (\vec {k})= 1$ (noise neglection) to reverse beam smoothing? 
No theoretical arguments have been found to answer this 
question. Numerical simulations have been necessary.
Results obtained from simulations are displayed in 
Fig. 4. In all the cases studied, the noise has been neglected and Eq.
(\ref{conv}) has been used to perform FTS--deconvolution.
Good results indicate that noise neglection is appropriate.

In the left panels of Fig. 4, the level of noise is $16.4 \ \mu K$. 
For $\theta_{_{FWHM}}=8.8^{\prime}$ (top left panel)
spectra $E_{1 \ell }$ and $E_{3 \ell }$ are quasi indistinguishable in the
$\ell $--interval (2,2000). For $\theta_{_{FWHM}} \sim 9.5^{\prime}$, 
a small difference between these spectra appears for 
$1000 \leq \ell \leq 2000$ (middle left panel). Finally,  
$E_{1 \ell }$ and $E_{3 \ell }$ are clearly different for 
$\theta_{_{FWHM}}=10^{\prime}$ (bottom left panel). Hence,  
the maximum value of $\theta_{_{FWHM}}/ \Delta$ is close to 2
(around 12 pixels inside the beam circle).
The right panels of Fig. 4 show the same analysis as
the left panels for a noise level of $54.6 \ \mu K$. In this case,
spectra $E_{1 \ell }$ and $E_{3 \ell }$ are similar in all the 
interval (2,2000) for $\theta_{_{FWHM}}<8^{\prime}$ (top right panel), 
small discrepancies in the interval (1000,2000) have already 
appeared for $\theta_{_{FWHM}} \sim 8.5^{\prime}$ 
(middle right panel) and, 
finally, for $\theta_{_{FWHM}} \sim 9^{\prime}$ 
these discrepancies are important (bottom right panel).
For this level of noise, the maximum value of $\theta_{_{FWHM}}/ \Delta$ 
is close to 1.8
(around 10 pixels inside the beam circle). 
This qualitative analysis of Fig. 4 is confirmed by 
Table 3 where quantities $M1$, $M2$, $MA$, and $\Sigma$
are presented in all the cases.

\section{PIXELIZATION
AND DECONVOLUTION IN REALISTIC EXPERIMENTS}

Limitations of the S--deconvolution process have been discussed
along the paper.
If the pixel size $\Delta$ is taken to be similar to the beam radius
$\theta_{_{FWHM}}$, the angular power spectrum can be
only estimated for $\ell \leq  \ell_{max}$ with
$\ell_{max} \sim 180/ \Delta \sim 180/ \theta_{_{FWHM}}$;
however, for $\Delta \sim \theta_{_{FWHM}}/2$, 
the angular power spectrum can be evaluated
up to $\ell_{max} \sim 360/ \theta_{_{FWHM}}$. 
In these formulae, $\Delta$ and $\theta_{_{FWHM}}$ must be written
in degrees.
We see that, for 
$\theta_{_{FWHM}} \sim 10^{\prime}$, the pixelization 
$\Delta \sim \theta_{_{FWHM}}$ ($\Delta \sim \theta_{_{FWHM}}/2$)
allows us to get the spectrum up to $\ell_{max} \sim 1080$
($\ell_{max} \sim 2160$); therefore, for a given beam, 
the choice of the best feasible pixelization is crucial
in order to get maximum information from observations.
The minimum pixel size compatible with S--deconvolution (for the 
methods used in the paper) is hereafter denoted $\Delta_{_{DE}}$.

In realistic experiments, various effects --apart from deconvolution--
conditionate the choice of the most appropriate  pixelization.
In order to discuss these effects,
let us focus our attention on PLANCK project (see Tauber, 1999).
With a telecope having  a diameter $D$, the minimum pixel size
allowed by {\em difraction} is roughly $\Delta_{_{DI}} \sim 1.22c/D \nu$, 
where 
$c$ is the speed of light and $\nu$ its frequency; 
hence, this minimum size depends 
on $\nu$. In a multifrequency experiment, 
there is a minimum pixel size $\Delta_{_{DI}}$ corresponding to
each frequency; for example, in the PLANCK mission ($D=1.5 \ m$), 
observations will be 
carried out in nine different frequencies ranging from $30 \ GHz$
to $857 \ GHz$ and, consequently, the size $\Delta_{_{DI}}$ ranges
from $ 30^{\prime} $ to $ 1^{\prime} $. 
Furthermore, 
in the PLANCK case,
the line of sight will move on a big circle in the sky each minute; 
hence, if
two successive temperature assignations on the circle are 
performed at an angular distance $\Delta \alpha $,
the time --in seconds-- between these asignations is 
$\Delta t = 2.78 \times 10^{-3} \Delta \alpha $.
The angle $\Delta \alpha $ must be chosen in such a way that
(1) no large overlaping of contiguous beam positions occurs and
(2) time $\Delta t$ is greater that the response time of the bolometers.
For the chosen period of one minute and $\Delta \alpha \geq
2 \theta_{_{FWHM}}$, there is no overlaping and
condition (2) is satisfied for the
PLANCK bolometers. 
A certain pixel size is only 
admissible if technology plus observational strategy ensure that 
each pixel 
is observed a large enough number of times during the mission.  
Let us estimate this number for PLANCK. 
For a pixel size $\Delta$, the total number of pixels is
$N = 1.5 \times 10^{8} \Delta^{-2}$ and admitting uniform coverage 
during a year
(for qualitative estimates) each pixel is observed for a time 
$\Delta t_{p} = 0.2 \Delta^{2} \ s$; therefore, the number of 
observations per pixel is $N_{p} = \Delta t_{p} / \Delta t =
72 \Delta^{2} / \Delta \alpha $. Finally, 
for $\Delta \alpha =
2 \theta_{_{FWHM}}$, one easily see 
that the size necessary to 
obtain $N_{p}$ observations by pixel (during a year of PLANCK mission)
is
$\Delta^{*} = \frac {1}{6} (N_{p} \theta_{_{FWHM}})^{1/2}$.
Since the $\theta_{_{FWHM}}$ values for PLANCK detectors 
range from $\sim 30^{\prime}$ to $\sim 4.5^{\prime}$, assuming
$N_{p} \geq 100$, we see that $\Delta^{*}$ 
ranges from $\sim 9.5^{\prime}$ 
to $\sim 1.3^{\prime}$. This means that, in order to 
have a number of observations by pixel greater than 100, the
pixel size must be greater than $\Delta^{*}(N_{p}=100)=\Delta_{100}$.

Given a frequency, there is an optimum pixel size, $\Delta_{_{OP}}$, 
which will be assumed to be  the maximum of the three above 
sizes $\Delta_{DE}$,
$\Delta_{DI}$ and $\Delta_{100}$.
The value of $\Delta_{_{OP}}$ depends on the frequency.

In order to separate the foregrounds and the cosmological signal 
in multifrequency experiments, various frequencies 
and a unique pixelization must be used and, consequently,
the best pixelization would be the maximum of the 
$\Delta_{_{OP}}$ optimal sizes corresponding to the
involved frequencies. This maximum corresponds to the lowest
frequency under consideration; for example, in the PLANCK case,
if all the frequencies from $30 \ GHz$ to $857 \ GHz$ are 
considered, the minimun admissible pixel appears to have a size
$\Delta = \Delta_{_{DI}} (\nu = 30) \sim 30^{\prime}$, for which,
the spectrum can be only estimated up to $\ell_{max} = 360$.
Of course, we could consider only the frequencies greater than 
$53 \ GHz$ (with some loss of information) and, then, the minimum pixel is 
$\Delta = \Delta_{_{DI}} (\nu = 53) \sim 17^{\prime}$ and
$l_{max} = 630$ and so on.

Let us reconsider the
radiometer 
working at $\nu \sim 100 \ GHz$ 
with $\theta_{_{FWHM}}=10^{\prime}$, which was projected to be 
inside PLANCK satellite. For this radiometer one easily find
$\Delta_{_{DE}} \simeq 5^{\prime}$,
$\Delta_{_{DI}} = 8.4^{\prime}$ and $\Delta_{100} \simeq 3^{\prime}$; hence, 
$\Delta_{_{OP}} = \Delta_{_{DI}} =8.4^{\prime}$ ($\ell_{max} \sim 1290$), 
for this optimum pixelization, the level of noise is $5.3 \times 10^{-6}$
and the angular spectrum can be obtained 
for $\ell \leq 1290$ (top panel of Fig. 5 and entries 1 and 2 of 
Table 4).
We now consider the      
bolometer
working at $\nu \sim 143 \ GHz$ 
with $\theta_{_{FWHM}}=10.3^{\prime}$, which
was also proposed to measure CMB anisotropy from the PLANCK
satellite (phase A study). For this detector we
easily find
$\Delta_{_{DE}} \simeq 5^{\prime}$,
$\Delta_{_{DI}} \simeq 5.87^{\prime}$ and $\Delta_{100} \simeq 5^{\prime} $; 
hence, 
$\Delta_{_{OP}} = \Delta_{_{DI}} =5.87^{\prime}$ ($\ell_{max} \sim 1840$).
For the pixelization $\Delta = 5.87^{\prime}$ 
the level of noise is $1.28 \times 10^{-5}$ and 
we have verified that 
FTS--deconvolution 
leads to the right spectrum for $\ell \leq 1840$ (bottom panel of Fig. 5
and entries 3 and 4 of Table 4).

For the bolometer working at $857 \ GHz$ with 
$\theta_{_{FWHM}}=4.4^{\prime}$, we get
$\Delta_{_{DI}} = 1^{\prime}$ and $\Delta_{100} \simeq 1^{\prime}$,
but the estimate of 
$\Delta_{DE}$ is problematic as a result of the high level of
noise of this bolometer. Perhaps, in this case,  
maximum entropy or wavelets could give good results; for example,
wavelets could be used to lower the noise before beam
deconvolution.

\section{DISCUSSION AND CONCLUSIONS}

We expect that beam deconvolution will be important in order to
study some aspects of the observational maps given by
experiments as PLANCK. As an example, let us argue that 
the study of the 
statistical properties of a given observational map
should be performed after deconvolution. In fact, 
various methods can be used to know if
the maps are Gaussian or they obey other statistics;
among them, the estimation of the correlation 
function of pixels where the signal is above 
a certain threshold (excursion sets, Kaiser 1984) and the local 
analysis of the 
spots distributed in the map (Bond and Efstathiou, 1987). 
Since the beam smoothes the map, 
it alters the correlations of excursion sets and the
structure and distribution of the spots; hence, 
the above 
methods for analyzing statistics 
should be applied after a good deconvolution.

The separation of the cosmic signal and the
foregrounds requires a unique appropriate pixelization.
In the PLANCK case, we have seen that the optimal size for
this pixelization,
$\Delta_{_{OP}}$, coincides with the size $\Delta_{_{DI}}
\sim \theta_{_{FWHM}}$     
corresponding to the lowest frequency under
consideration; nevertheless, other studies 
can be imagined (statistical analysis et cetera) which could be
performed on the maps corresponding to a given 
frequency (without previous separation). 

For small enough (but feasible) values of $\sigma_{_{N}}$ plus 
a certain beam (either spherical with a $\theta_{_{FWHM}}$ or asymmteric),
our codes allow us to find the most appropriate 
pixel size for deconvolution $\Delta = \Delta_{_{DE}}$.
For the corresponding pixelization,
FTS--deconvolution leads to 
a good estimation of the angular power
spectrum in the most wide $\ell$--interval.
The size $\Delta_{_{DE}}$ must be compared to
$\Delta_{100}$ and $\Delta_{_{DI}}$ to choose the most
appropriate pixelization $\Delta = \Delta_{_{OP}}$. 
In the case $\Delta_{_{OP}} > \Delta_{_{DE}}$, 
the study about beam reversion presented in 
\S \ 3 to \S \ 5 proves that                           
S--deconvolution can be performed
using very simple methods.
For levels of noise much higher that those of previous 
sections, further study is necessary; maximum entropy,
wavelets or other methods should be tried out.

The goodness of a certain pixelization against beam 
S-deconvolution has appeared to be weakly dependent on the particular
mathematical 
method used to reverse the beam average. This fact suggests
that deconvolution procedures different from those
of this paper could alter its results. 
Altough this suggestion should be a 
motivation for studying new methods to get approximate
solutions of Eqs. (6) (S--deconvolutions), 
the structure of the system 
of linear equations to be solved is always the same
and, consequently, all the mathematical methods
could exhibit similar limitations to solve it.
Indeed, we believe that new deconvolution
methods could lead to some
modifications of the results of 
this paper, but not to very different
values of $\Delta_{_{DE}}$.  

Our estimates show that, in the absence of any problem with beam 
asymmetry,
current technology could lead to a good estimate of the  
angular power spectrum (of the total signal including foregrounds)
in a $\ell$--interval which depends 
on frequency. In the particular case of two instruments
on board of PLANCK,
we have used optimum pixelization and forty
$20^{\circ} \times 20^{\circ}$ maps to find that the
spectra are recovered from 
$\ell=200$ to $\ell \sim 1300$ in the case of a radiometer and
from 
$\ell=200$ to $\ell \sim 1800$ for a certain bolometer.
Since the sky can be divided into $\sim 100$
of these maps and we only need 
about $40$ for a good estimate of the spectrum (for large $\ell$
values), we can select the best forty maps; namely, the maps having
minimum contaminations. The uncertainty produced by this partial coverage
appears to be a little greater than the errors produced by the 
implemented 
deconvolution procedures (this means that these procedures
are good enough for us).

The problem with the deviations of the beam structure with
respect to spherical symmetry deserves much attention.
As discussed above, FTS--deconvolution is compatible with
beam asymmetry if the beam orientation is preserved
from measurement to measurement. If the experiment 
is designed in such a way that the beam does not rotate,
our codes for FTS--deconvolution work (see \S \ 3.4.2 and the 
middle and bottom right
panels of Fig. 2); 
however, 
if the beam rotates,
FTS--deconvolution does not apply and, moreover,
operative methods for making beam S--deconvolution are not known; 
hence,
if the assymetry is high enough, 
beam S--deconvolution is not feasible (so far).
In short, 
excepting the case of negligible deviations with respect to 
spherical symmetry,
{\em any effort directed to maintain unaltered the beam orientation
during observations seems to be of great interest}. Unfortunately,  
in spatial projects as Planck, the design of the
observational strategy
does not preserve this orientation. 

Given an asymmetric rotating beam, it would be interesting to study
the whole effect of asymmetry plus changing orientation. 
Even if 
S--deconvolution is not feasible, the estimation of this whole effect 
could be a further direct 
application of the techinques used in this paper.
The following method 
seems to be appropriate: (i) average the asymmetric beam
--on appropriate shells--  to get a new associated one 
with spherical symmetry,
(ii) simulate forty S1 maps, (iii) smooth the S1 maps {\em with the
assymmetric beam} taking into account the orientation change 
produced by the observational strategy; thus, we obtain
forty S2 maps, (iv) deconvolve the S2 maps {\em with the spherically
symmetric beam of reference} to get the S3 maps, and (v)
estimate the spectra $E_{1 \ell}$, $E_{2 \ell}$ and, $E_{3 \ell}$.
If $E_{1 \ell}$ and $E_{3 \ell}$ are similar enough, the assymetry
can be neglected, on the contrary, the differences between these two spectra 
can be considered as a measure of the whole effect
of asymmetry plus rotation. This study should be developed for realistic
beams and observational strategies, 
which is out of the scope of this paper.

\newpage

\noindent
{\bf ACKNOWLEDGMENTS}. This work has been partially
supported by the Spanish DGES (project PB96-0797). Some calculations
were carried out on a SGI Origin 2000s at the Centro de Inform\'atica
de la Universidad de Valencia.

\vspace {0.7 cm}

\noindent
{\large{\bf REFERENCES}}\\
\\
Bond, J.R., \& Efstathiou, G., 1987, MNRAS, 226, 655\\
Golub, G.H., \& van Loan, C.F., 1989, Matrix Computations
(Baltimore: Johns Hopkins Press)\\
G\'{o}rski, K.M., Banday, A.J., Bennett, C.L., Hinshaw, G.,
Kogut, A., Smoot, G.F., \& Wright, E.L. 1996, ApJ, 464, L11\\
Kaiser, N., 1984, ApJ, 284, L9\\
Knox, L., 1995, Phys. Rev., 52D, 4307\\
Press, W.H., Flannery, B.P., Teukolski, S.A., \&
Vetterling, W.T., 1988, Numerical Recipes (New York: Cambridge
University press)\\
S\'aez, D., \& Arnau, J.V., 1997, ApJ, 476, 1\\
S\'aez, D., Holtmann, E., \& Smoot, G.F., 1996, ApJ, 473, 1\\  
Young, D.M., 1971, Iterative Solution of Large Linear Systems
(New York: Academic Press)\\
Sugiyama, N., 1995, ApJ Supplement, 100, 281\\
Tauber, J.A., 1999, Astrophys. Lett. Comm., in press
\newpage

\begin{center}
{\bf FIGURE CAPTIONS}
\end{center}

\vskip 0.5cm
\noindent

\noindent
{\bf FIG.\ 1.--} Top (middle) panel shows the relation between 
the size of a circular beam and that of the
pixels for the most dense pixelization 
compatible with LS (FTS) deconvolution. Bottom panel shows the
same for an asymmetric beam and FTS--deconvolution

\vskip 0.5cm

\noindent
{\bf FIG.\ 2.--} 
Each panel shows 
quantity $\ell (\ell+1) E_{\ell} \times 10^{10}$
versus $\log \ell$. No noise is present and $\Delta = 4.6875^{\prime}$.
All panels, excepting the bottom right one, contain three lines:
pointed line gives the spectrum 
before beam smoothing, dashed line is the spectrum after 
smoothing and, solid line corresponds to the S--deconvolved 
spectrum. Left: top, middle and bottom panels
correpond to FTS--deconvolutions with beams having
$\theta_{_{FWHM}}=10^{\prime}$, 
$\theta_{_{FWHM}}=11^{\prime}$, and
$\theta_{_{FWHM}}=11.5^{\prime}$, respectively.
Right: top (middle) panel shows the same as the left 
panels for LS--deconvolution and $\theta_{_{FWHM}}=8.8^{\prime}$
(for FTS--deconvolution and the asymmetric
beam defined in the text). 
The dotted (solid) line of the bottom right panel gives the
spectrum after smoothing for a spherically symmetric beam 
with $\sigma = 10^{\prime}$ (for the asymmetric beam of the text).

\vskip 0.5cm 

\noindent
{\bf FIG.\ 3.--} The quantities represented are the same 
as in all the panels of Fig. 2. Solid (pointed) line is 
the modified spectrum extracted from a coverage of forty
$20^{\circ} \times 20^{\circ}$ maps (the theoretical
modified spectrum). 

\newpage
\noindent
{\bf FIG.\ 4.--} The same as in the left panels of Fig. 2.
Left: the
level of uncorrelated noise per pixel is $\sigma_{_{N}} = 16.4 \ \mu K$. 
Top, middle and bottom panels
correpond to $\theta_{_{FWHM}}=8.8^{\prime}$, 
$\theta_{_{FWHM}}=9.5^{\prime}$,  and
$\theta_{_{FWHM}}=10^{\prime}$, respectively.
Right: the same as in left panels for $\sigma_{_{N}} = 54.6 \ \mu K$.
Top, middle and bottom panels
correpond to $\theta_{_{FWHM}}=8^{\prime}$, 
$\theta_{_{FWHM}}=8.5^{\prime}$,  and
$\theta_{_{FWHM}}=9^{\prime}$, respectively. Pixel size is
$\Delta = 4.6875^{\prime}$ in all cases.

\vskip 0.5cm 

\noindent
{\bf FIG.\ 5.--} The same as in Fig. 4.
Top panel shows the results of FTS--deconvolution for a radiometer 
of the PLANCK
mission described in the text. The pixel size is  
$\Delta=8.4^{\prime}$. The
bottom panel shows the same for
a bolometer of PLANCK and $\Delta=5.87^{\prime}$.

\newpage 

\begin{table}
\begin{center}
{\bf TABLE 1}\\ 
CONVERGENCE OF THE\\
JACOBI METHOD\\
\begin{tabular}{cccc}\\
\hline
\hline 
$q$ & $\lambda_{max}$ & $(\theta_{_{FWHM}}/ \Delta)_{max}$ 
& $S_{_{q}}/S_{_{B}}$ \\
\hline
3 & 0.5000 & 1.998 & 0.72\\ 
5 & 0.4565 & 1.879 & 2.25\\      
7 & 0.4559 & 1.877  & 4.43\\
9. &0.4559 & 1.877 & 7.32\\
\hline
\multicolumn{4}{c}{}\\
\end{tabular}
\end{center}   
\end{table} 

\newpage 

\begin{table}
\begin{center}
{\bf TABLE 2}\\ 
COMPARING SPECTRA\\ 
FIGS. 2 AND 3\\  
\begin{tabular}{cccccc}\\
\hline
\hline
Panel & $\ell$--Interval & $M1$ & $M2$ & $MA$ &  $\Sigma$ \\
\hline
Fig2, Top-Left & 40--1000 & 13.22 & $4.94 \times 10^{-2}$  & 0.23 & 0.30\\ 
Fig2, Top-Left & 1000--2000 & -14.97 & $-1.25 \times 10^{-2}$  & 0.25 & 0.32\\
Fig2, Middle-Left & 40--1000 & 13.14 & -0.12 & 0.27 & 0.35\\ 
Fig2, Middle-Left & 1000--2000 & -14.90 & 0.79 & 0.90 & 1.08\\ 
Fig2, Bottom-Left & 40--1000 & 13.21 & -1.12 & 1.12 & 1.21\\
Fig2, Bottom-Left & 1000--2000 & -14.96 & 4.43 & 4.52 & 5.51\\
Fig2, Top-Right & 40--1000 & 13.28 & $-5.68 \times 10^{-3}$ & 0.18 & 0.21\\ 
Fig2, Top-Right & 1000--2000 & -15.04 & $8.95 \times 10^{-3}$ & 0.27 & 0.34\\ 
Fig2, Middle-Right & 40--1000 & 13.24 & $8.88 \times 10^{-2}$ & 0.22 & 0.27\\
Fig2, Middle-Right & 1000--2000 & -14.99 & 
$-9.76 \times 10^{-2}$ & 0.36 & 0.45\\
Fig2, Bottom-Right & 40--1000 & 9.79 & -3.38 & 3.47 & 4.36\\
Fig2, Bottom-Right & 1000--2000 & -11.65 & 3.32 & 3.65 & 4.03\\
Fig3 & 40--1000 & 13.48 & -0.35 & 0.61 & 0.76\\
Fig3 & 1000--2000 & -14.60 & $7.20 \times 10^{-2}$ & 0.39 & 0.48\\
\hline
\multicolumn{6}{c}{}\\
\end{tabular}
\end{center}   
\end{table}   

\newpage

\begin{table}
\begin{center}
{\bf TABLE 3}\\ 
COMPARING SPECTRA\\ 
FIG. 4\\  
\begin{tabular}{cccccc}\\
\hline
\hline
Panel & $\ell$--Interval & $M1$ & $M2$ & $MA$ &  $\Sigma$ \\
\hline
Top-Left & 40--1000 & 13.27 & $7.79 \times 10^{-2}$  & 0.20 & 0.23\\ 
Top-Left & 1000--2000 & -15.03 & 0.12  & 0.33 & 0.39\\
Middle-Left & 40--1000 & 13.20 & -0.17 & 0.24 & 0.30\\ 
Middle-Left & 1000--2000 & -14.96 & 1.10 & 1.17 & 1.45\\ 
Bottom-Left & 40--1000 & 13.30 & -1.13 & 1.13 & 1.26\\
Bottom-Left & 1000--2000 & -15.09 & 4.10 & 4.24 & 5.39\\
Top-Right & 40--1000 & 13.26 & 0.16 & 0.26 & 0.31\\ 
Top-Right & 1000--2000 & -15.02 & 0.34 & 0.39 & 0.54\\ 
Middle-Right & 40--1000 & 13.24 & $5.54 \times 10^{-2}$ & 0.22 & 0.29\\
Middle-Right & 1000--2000 & -14.99 & 1.24 & 1.28 & 1.47\\
Bottom-Right & 40--1000 & 13.27 & -0.43 & 0.51 & 0.64\\
Bottom-Right & 1000--2000 & -15.03 & 4.03 & 4.10 & 4.82\\
\hline
\multicolumn{6}{c}{}\\
\end{tabular}
\end{center}
\end{table} 

\newpage

\begin{table}
\begin{center}
{\bf TABLE 4}\\ 
COMPARING SPECTRA\\ 
FIG. 5\\
\begin{tabular}{cccccc}\\
\hline
\hline
Panel & $\ell$--Interval & $M1$ & $M2$ & $MA$ &  $\Sigma$ \\
\hline
Top & 40--700 & 8.96 & $0.12$  & 0.19 & 0.22\\ 
Top & 700--1290 & -14.99 & -0.22  & 0.27 & 0.33\\
Bottom & 40--1000 & 9.43 & 0.13 & 0.23 & 0.30\\
Bottom & 1000--1840 & -16.24 & 0.31 & 0.40 & 0.49\\
\hline
\multicolumn{6}{c}{}\\
\end{tabular}
\end{center}
\end{table} 

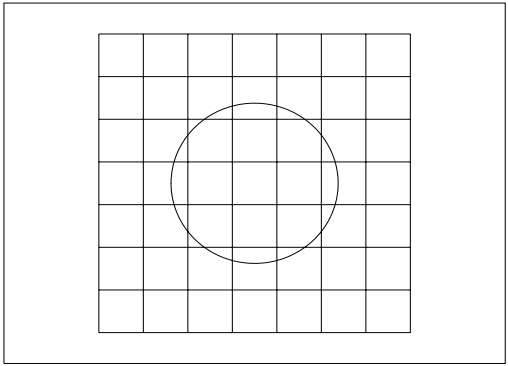
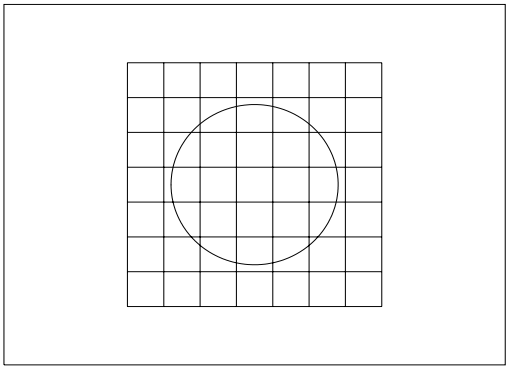
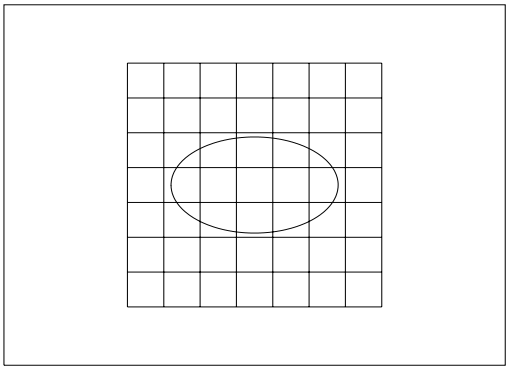
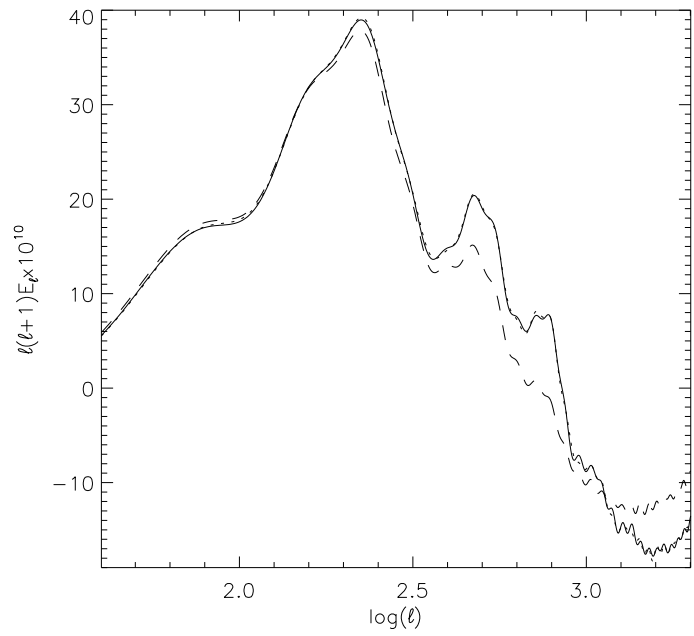
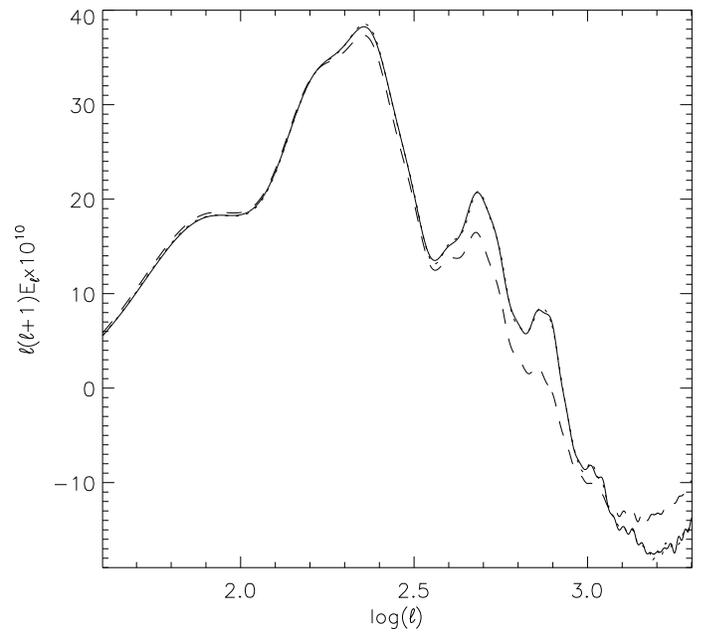
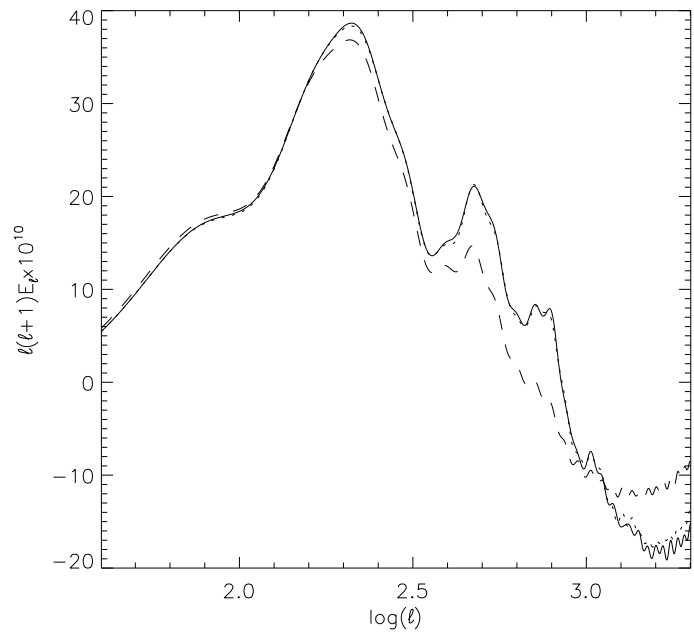
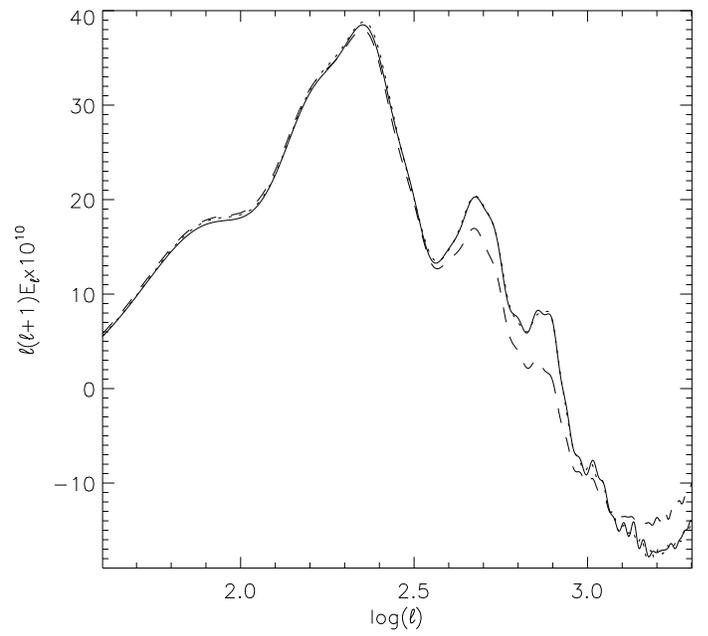
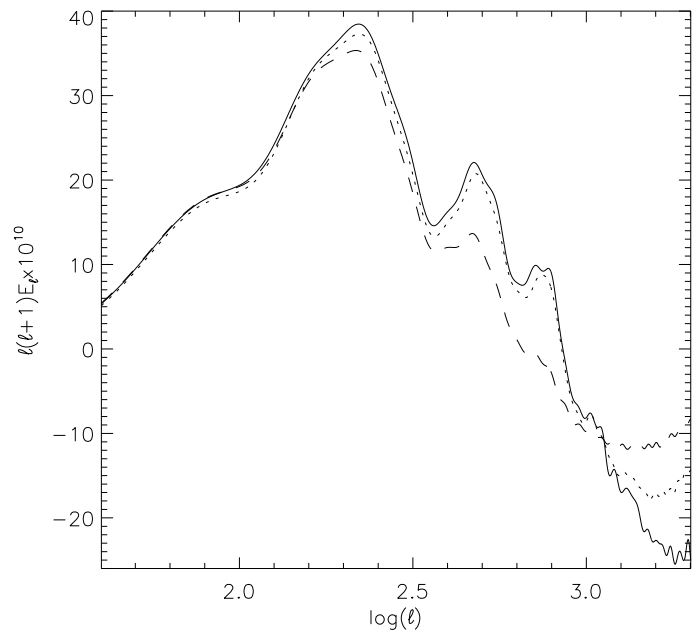
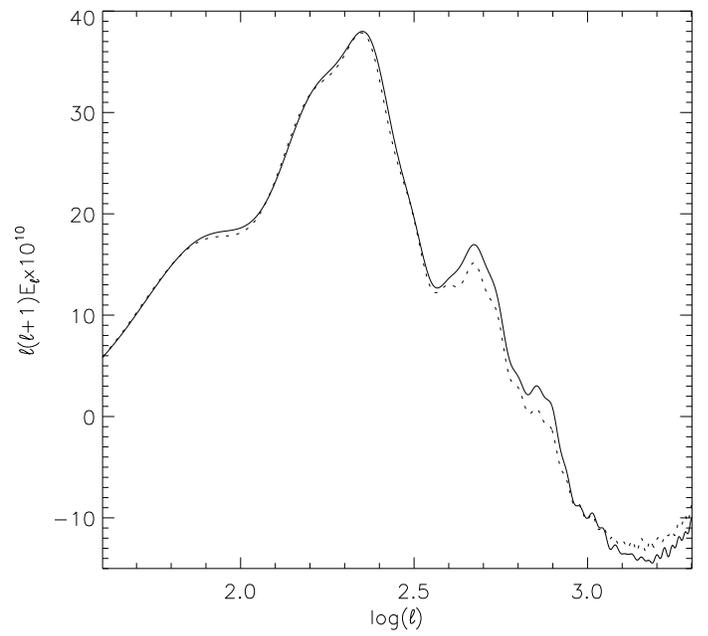
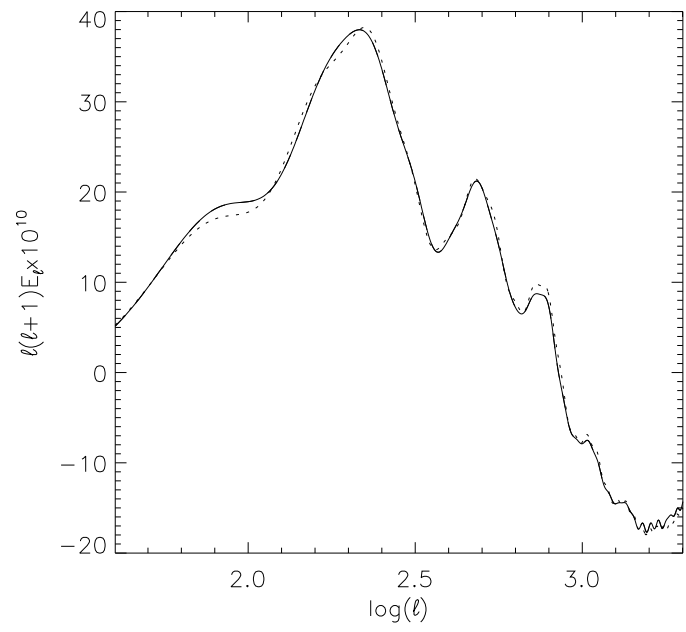
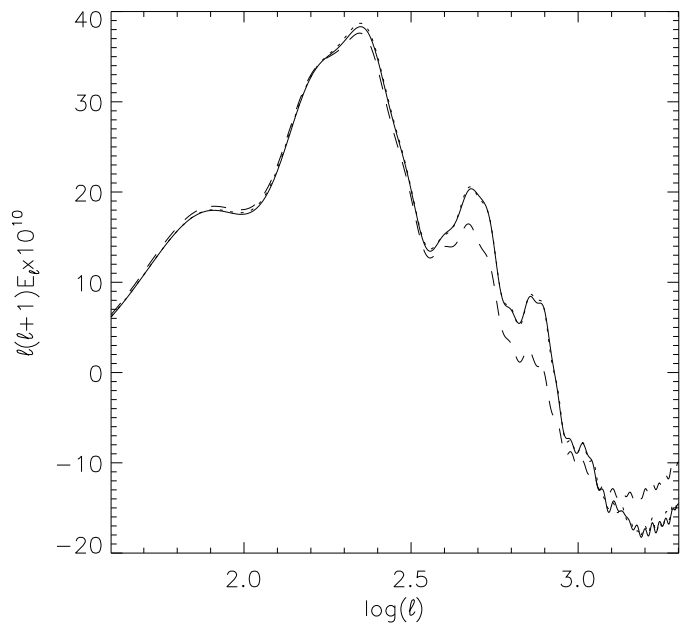
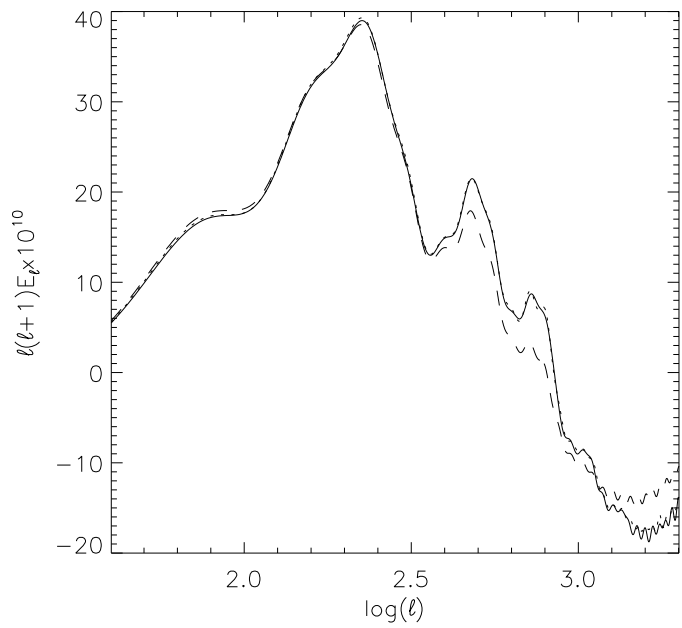
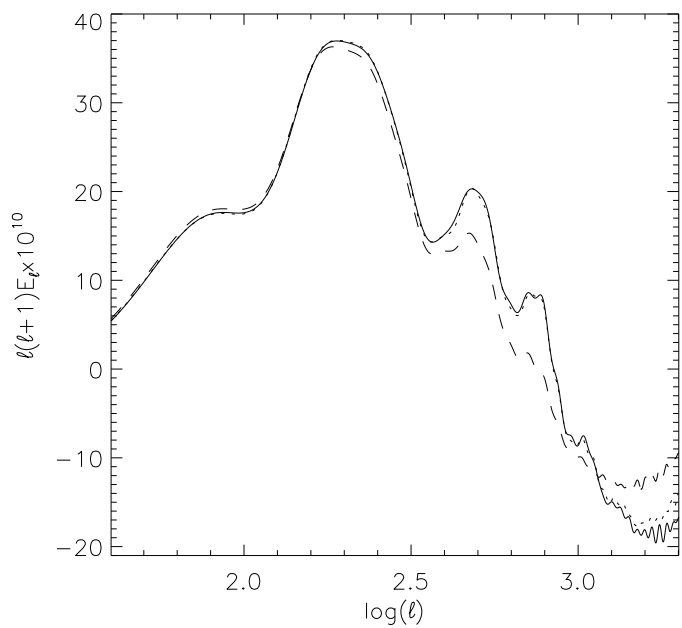
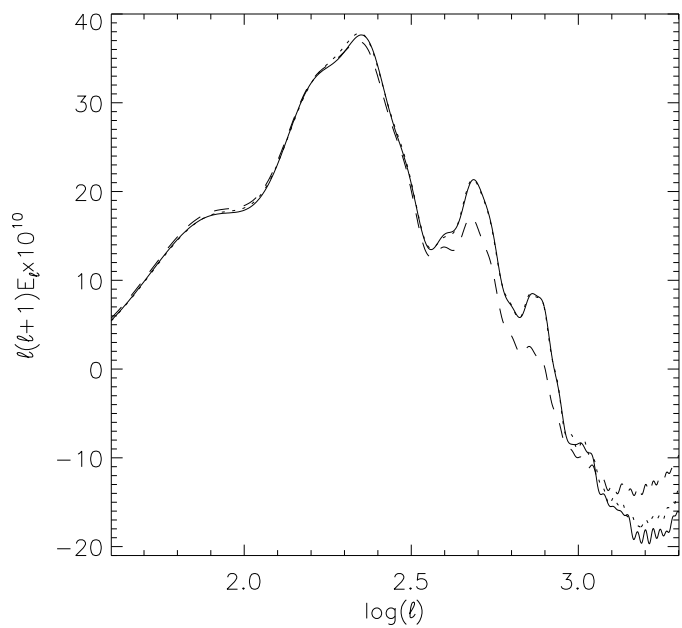
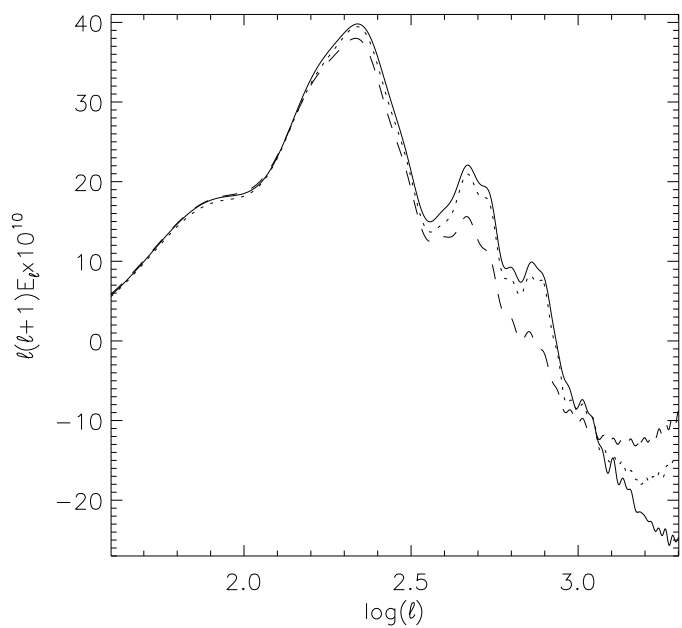
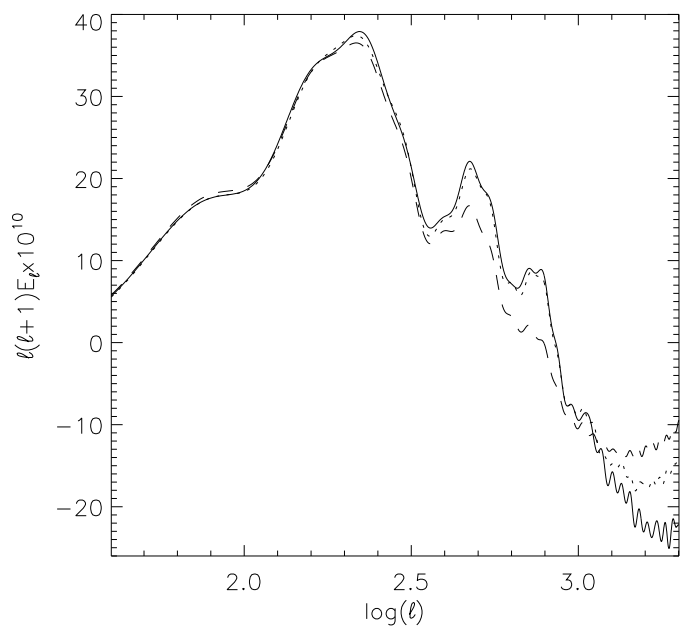
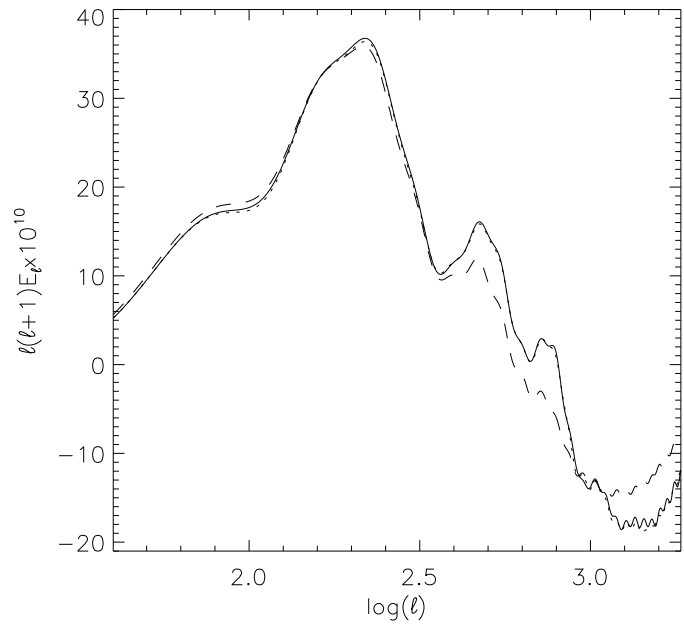
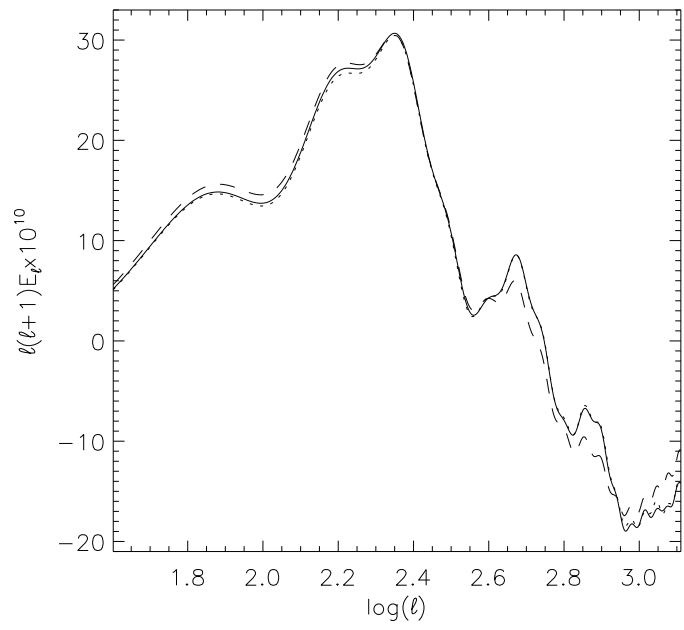
\end{document}